\pdfoutput=1
\documentclass[twocolappendix]{emulateapj}
\usepackage{amsmath}
\usepackage{graphicx,epstopdf}
\usepackage{multirow}
\usepackage{color}
\usepackage{latexsym}
\usepackage{amssymb}
\usepackage{epsfig}
\usepackage{verbatim}
\usepackage{placeins}

\usepackage[colorlinks=true,linkcolor=blue,citecolor=blue]{hyperref}

\begin{document}

\title{Blackbody Radiation from Isolated Neptunes}

\author{Sivan Ginzburg\altaffilmark{1}, Re'em Sari\altaffilmark{1,2,3}, Abraham Loeb\altaffilmark{3}}

\altaffiltext{1}{Racah Institute of Physics, The Hebrew University, Jerusalem 91904, Israel}

\altaffiltext{2}{Radcliffe Institute for Advanced Study, Harvard University, Cambridge, MA 02138, USA}

\altaffiltext{3}{Astronomy department, Harvard University, 60 Garden Street, Cambridge, MA 02138, USA}

\begin{abstract}
Recent analyses of the orbits of some Kuiper Belt objects hypothesize the presence of an undiscovered Neptune-size planet at a very large separation from the Sun. The energy budget of Neptunes on such distant orbits is dominated by the internal heat released by their cooling rather than solar irradiation (making them effectively ``isolated''). The blackbody radiation that these planets emit as they cool may provide the means for their detection. Here we use an analytical toy model to study the cooling and radiation of isolated Neptunes. This model can translate a detection (or a null detection) to a constraint on the size and composition of the hypothesized ``Planet Nine''.  Specifically, the thick gas atmosphere of Neptune-like planets serves as an insulating blanket which slows down their cooling. Therefore, a measurement of the blackbody temperature, $T_{\rm eff}\sim 50\textrm{ K}$, at which a Neptune emits can be used to estimate the mass of its atmosphere, $M_{\rm atm}$. Explicitly, we find the relation $T_{\rm eff}\propto M_{\rm atm}^{1/12}$. Despite this weak relation, a measurement of the flux at the Wien tail can constrain the atmospheric mass, at least to within a factor of a few, and provide useful limits to possible formation scenarios of these planets. 
Finally, we constrain the size and composition of Planet Nine by combining our model with the null results of recent all-sky surveys.
\end{abstract}
  

\section{Introduction}
The presence of an outer undetected planet in the solar system has been recently suggested, due to possible evidence of its gravitational influence on the orbits of Kuiper Belt objects \citep{Marcos2014,TrujilloSheppard2014,BatyginBrown2016}.
Here we focus on the scenario of a $\sim 10 M_{\Earth}$ planet on a $\sim 700 \textrm{ AU}$ semimajor axis orbit (as we show below, the distance from the Sun is not important for the planet's evolution, as long as it is large enough), as proposed by \citet{BatyginBrown2016} \citep[see also][]{Malhorta2016}. As such a planet cools, the internal flux it releases overwhelms the incident solar irradiation, and its thermal evolution is independent of the Sun, as if it were isolated. In particular, it has been argued that the internal luminosity released by the planet may allow its detection \citep{Cowan2016,LinderMordasini2016}.

In this paper, we study the cooling of isolated planets assuming a Neptune-like composition. Specifically, we relate the observable surface temperature and luminosity to the mass and composition of the planet. Special attention is given to the gas atmosphere of the planet, which governs its cooling rate. 

Unlike \citet{LinderMordasini2016}, we do not assume any specific formation scenario, and we do not couple the atmosphere mass fraction to the planet's mass. Rather, we analytically derive scaling laws connecting the luminosity to the mass of the planet and to the mass of the gas envelope.

The outline of the paper is as follows. In Section \ref{sec:two_layer} we present our two layer model for Neptune-like planets and in Section \ref{sec:cooling} we calculate its evolution over time. Section \ref{sec:composition} relates the observable radiation to the planet's size and composition, and derives associated constraints from recent all-sky surveys. Our conclusions are summarized in Section \ref{sec:conclusions}.

\section{Two Layer Neptune Model}\label{sec:two_layer}

Below we present a simplified model for Neptune-like planets, which contains the essential ingredients needed to describe their thermal evolution.

We model the planet as having a rocky and icy core, which amounts to most of the planet's mass $M$ and radius $R$. The core is surrounded by an atmosphere with a mass $M_{\rm atm}<M$ and a thickness $\Delta R<R$. The assumption of an envelope mass fraction which is smaller than unity $f\equiv M_{\rm atm}/M\lesssim 0.5$ is natural in the context of the core-nucleated accretion theory of giant planet formation \citep{PerriCameron1974,Harris1978,Mizuno1978,Mizuno1980,Stevenson1982}. According to this theory, once a rocky core accretes roughly its own mass in gas from the protoplanetary gas-rich disk, a runaway accretion initiates and the planet quickly evolves into a gas giant with mass $M\gg 10M_{\Earth}$ \citep{BodenheimerPollack86,Pollack96,PisoYoudin2014,Piso2015}.

We approximate the core as an incompressible fluid with a constant density. As we show below, the overlying atmosphere dictates a high temperature at the core-envelope boundary, inhibiting the formation of a solid crust, and keeping the entire core molten. We therefore model the core as convective (i.e. having uniform entropy), and since it is approximated as incompressible, it is also isothermal. These simplifications are in accordance with more sophisticated numerical models of Uranus and Neptune, in which the density and temperature of the core vary only by a factor of a few, while the pressure changes by orders of magnitude \citep{HubbardMacfarlane1980,GuillotGautier2014}.

\subsection{Gas Atmosphere}\label{sec:atmosphere}

We model the atmosphere as an ideal hydrogen and helium gas with a polytropic equation of state $P\propto\rho^{\gamma}$, with $P$ and $\rho$ denoting the pressure and density respectively and $\gamma$ is the polytropic index. Hydrostatic equilibrium leads to the following temperature profile as a function of depth $x$ inside the atmosphere
\begin{equation}\label{eq:tmp_prof}
k_{\rm B}T(x)=\frac{\gamma-1}{\gamma}\mu_g gx,
\end{equation}
where $k_{\rm B}$ is the Boltzmann constant, $\mu_g$ is the mean molecular mass of the gas and $g=GM/R^2$ is the surface gravity (with $G$ being Newton's constant).
The equilibrium temperature that the solar irradiation dictates at the outer boundary  (in the absence of an internal heat source), $T_{\rm eq}\approx 10\textrm{ K}$, is negligible, as we show below. Equation \eqref{eq:tmp_prof} relates the temperature at the base of the atmosphere, which is also the temperature of the adjacent core, to the thickness of the atmosphere  
\begin{equation}\label{eq:tmp_base_neptune}
T_c=1.7\times 10^4\textrm{ K}\left(\frac{M}{M_{\rm N}}\right)^{3/4}\left(\frac{\Delta R}{R}\right),
\end{equation}
where we scale to Neptune's mass, $M_{\rm N}$, by taking into account the slight gravitational compression of the core and using the relation $R\propto M^{1/4}$ instead of a constant density  relation $R\propto M^{1/3}$ \citep[e.g.,][]{Valencia2006,Baraffe2014}.

For present-day Neptune, Equation \eqref{eq:tmp_base_neptune} predicts $T_c\approx 3400\textrm{ K}$ for $\Delta R/R\approx 0.2$, similar to numerically calculated structures \citep{HubbardMacfarlane1980}, and ensuring that the core is molten. We note that the deep interior of a rocky (or iron) core might solidify due to the high pressures, but we still model it as isothermal \citep[see also][]{HubbardMacfarlane1980}. Icy cores, on the other hand, do not solidify at these temperatures \citep{Redmer2011}.

\section{Cooling and Evolution}\label{sec:cooling}

In a radiative envelope, the diffusion approximation sets a temperature profile, $T\propto\tau^{1/4}$, as a function of the optical depth $\tau$ (see the \hyperref[sec:convective]{Appendix}, which also discusses convective profiles). The effective surface temperature $T_{\rm eff}$ which sets the luminosity is determined at $\tau\sim 1$ and is given by
\begin{equation}\label{eq:tmp_ratio}
\begin{split}
\frac{T_c}{T_{\rm eff}}&=\tau_{\rm atm}^{1/4}\equiv\left(\frac{\kappa M_{\rm atm}}{4\pi R^2}\right)^{1/4}\\&\approx 53\left(\frac{R}{R_{\rm N}}\right)^{-1/2}\left(\frac{M_{\rm atm}}{M_\Earth}\right)^{1/4},
\end{split}
\end{equation}
with $\tau_{\rm atm}\equiv\int{\kappa\rho\rm{d}x}$ estimated by assuming a constant opacity $\kappa\sim0.1\textrm{ cm}^2\textrm{ g}^{-1}$, which is a reasonable approximation for the low effective temperatures we find below \citep{Alexander1989,BellLin1994}. A variable opacity is discussed in the \hyperref[sec:convective]{Appendix}. $R_{\rm N}$ is the radius of Neptune. 
Combining the results of Section \ref{sec:atmosphere} (specifically, $T_c\approx 3400\textrm{ K}$) with Equation \eqref{eq:tmp_ratio} implies an effective temperature of $T_{\rm eff}\approx 60\textrm{ K}$ for present-day Neptune, assuming a mass fraction $f\sim 10\%$ \citep[see, e.g.,][]{HubbardMacfarlane1980,Guillot1999}, consistent with observations \citep{Hildebrand1985} and numerical models \citep{FortneyNettelmann2010}.
 
Neptune's effective temperature is higher than its equilibrium temperature $T_{\rm eq}=47\textrm{ K}$ by a factor of 1.27, implying that its internal flux exceeds the incident solar irradiation by a factor of 1.6 \citep{PearlConrath1991}. While Neptune is only marginally isolated (the internal heat is comparable to the solar irradiation), we focus here on lower equilibrium temperatures $T_{\rm eq}\approx 10\textrm{ K}$, for which the planet can be treated as isolated. 

We now proceed and formulate a time-dependent cooling model. Such a model is made simple by noting that the ratio between the internal and effective temperatures $T_c/T_{\rm eff}$ remains roughly constant during the planet's evolution, as seen from Equation \eqref{eq:tmp_ratio}, following that the mass of the atmosphere $M_{\rm atm}$ is conserved, the radius is approximately constant because we assume $\Delta R<R$ \citep[this approximation improves as the atmosphere cools and shrinks with time; see Section \ref{sec:atmosphere} and][]{LinderMordasini2016} and we approximate the opacity as constant.

\begin{figure}[tbhp]
	\includegraphics[width=\columnwidth]{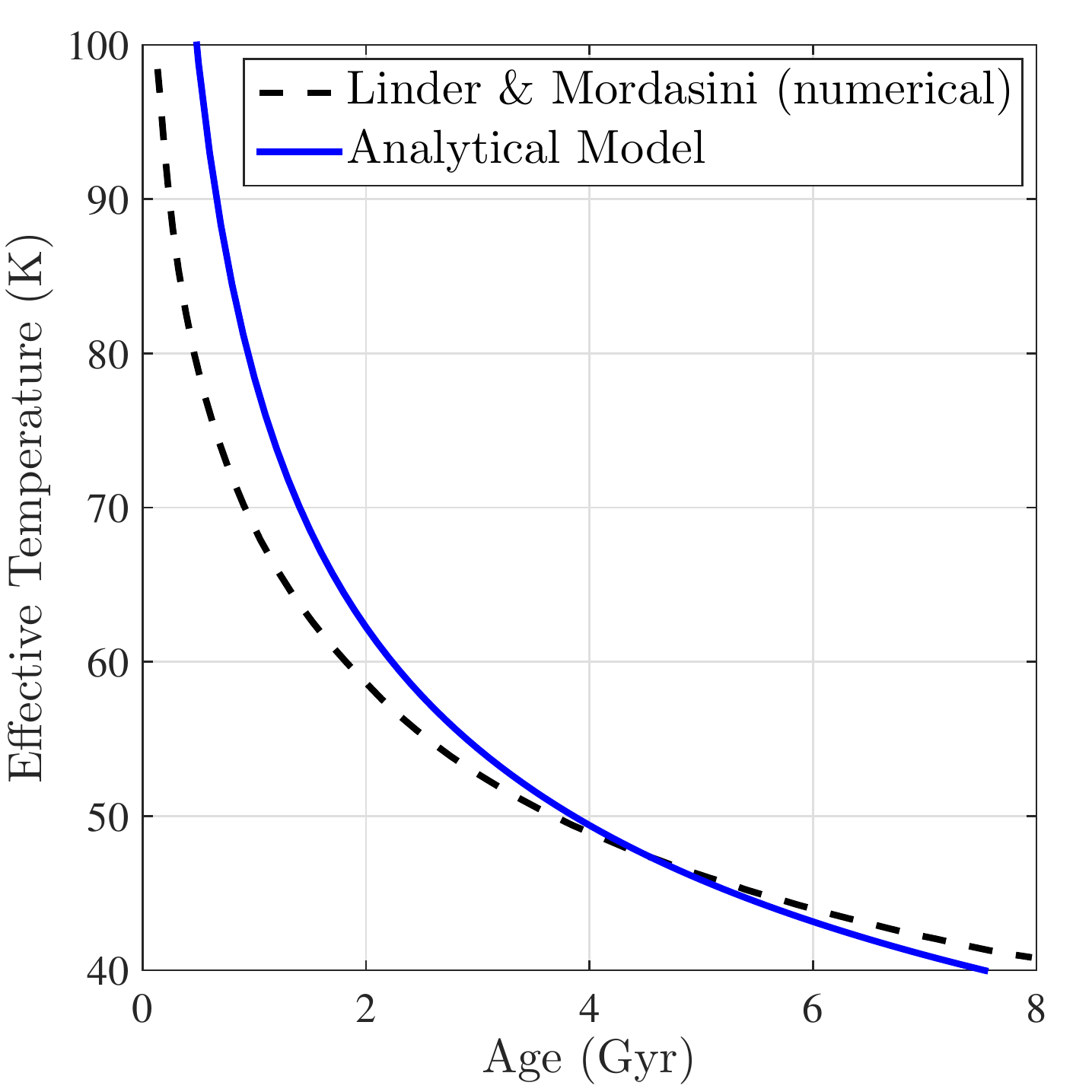}
	\caption{The effective temperature as a function of age for a $10M_\Earth=0.58M_{\rm N}$ planet with 14\% atmosphere by mass. The numerical results (dashed black line) are taken from \citet{LinderMordasini2016} while the analytical model (solid blue line) is according to Equation \eqref{eq:tmp_rad} and using $R\propto M^{1/4}$. 
		\label{fig:cooling}}
\end{figure}

The luminosity, i.e. cooling rate, is given by $L=4\pi R^2\sigma T_{\rm eff}^4$, with $\sigma$ denoting the Stephan-Boltzmann constant. The energy is given approximately by the Dulong-Petit law $E=3(M/\mu)k_{\rm B}T_c$, where $\mu$ is the mean molecular mass of the planet. By writing an evolution equation $L=-\dot{E}$, we find the effective temperature of the planet as a function of age $t$:
\begin{equation}\label{eq:teff_time}
T_{\rm eff}=\left(\frac{k_{\rm B}}{4\pi\sigma}\frac{M}{\mu}\frac{1}{R^2t}\right)^{1/3}\tau_{\rm atm}^{1/12}.
\end{equation}
Using Equation \eqref{eq:tmp_ratio}, we scale Equation \eqref{eq:teff_time} to our nominal model and to the age of the solar system:
\begin{equation}\label{eq:tmp_rad}
T_{\rm eff}\approx 50\textrm{ K}\left(\frac{R}{R_{\rm N}}\right)^{1/2}\left(\frac{M_{\rm atm}}{M_\Earth}\right)^{1/12}\left(\frac{t}{4.5\textrm{ Gyr}}\right)^{-1/3},
\end{equation}
where we multiply our initial result of 40 K by a fitting factor of 1.3 to match the internal heat of Neptune which corresponds to a temperature of $(T_{\rm eff}^4-T_{\rm eq}^4)^{1/4}\approx 50\textrm{ K}$.

In Figure \ref{fig:cooling} we compare the cooling of our toy model to numerical calculations by \citet{LinderMordasini2016}. While our simple model is not exact (specifically, $T_{\rm eff}\propto t^{-0.27}$ fits the numerical calculations better than our $T_{\rm eff}\propto t^{-1/3}$), the deviations are modest enough so that the model can provide approximate scaling laws relating the observed temperature and luminosity of an isolated Planet Nine to its mass and composition.

\section{Relating Observables to Composition}\label{sec:composition}

In this section we discuss the relation between the observable blackbody radiation from ``Planet Nine'' to its size and composition.

\citet{LinderMordasini2016} couple the mass fraction of the atmosphere to the mass of the planet, using planet formation simulations \citep{Mordasini2014}, relating larger values of $f$ to more massive planets, thus leaving only one parameter. Here we prefer not to make any prior assumptions on the formation scenario, so we treat the size and the atmosphere mass of the planet as two independent parameters. 

If we assume a specific composition, especially of the core (which dominates both the mass and radius), then the observed blackbody radius $R$ can be translated into the planet's mass $M$. Since the cooling of the planet is mediated by its gaseous atmosphere, as explained in Section \ref{sec:cooling}, the observed temperature can be used to estimate the mass of the atmosphere, as explicitly demonstrated by Equation \eqref{eq:tmp_rad}.  

\begin{figure}[tbhp]
	\includegraphics[width=\columnwidth]{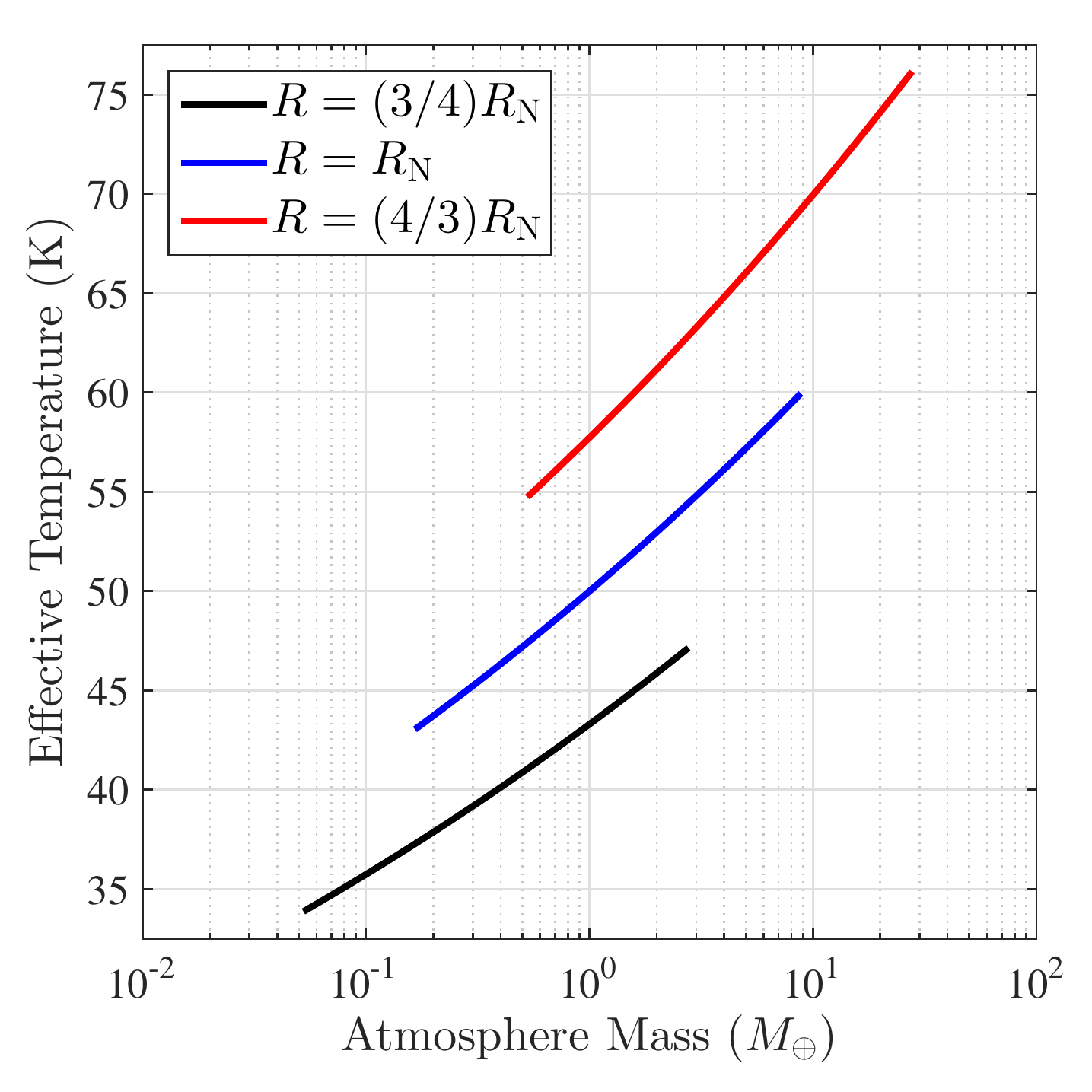}
	\caption{The effective temperature of 4.5 Gyr old planets as a function of their atmosphere's mass $M_{\rm atm}$. The curves follow Equation \eqref{eq:tmp_rad} for three values of the planet's radius $R$ in units of Neptune's radius: 3/4 (bottom black line), 1 (middle blue line), and 4/3 (top red line). These radii roughly correspond to planet masses of $5.4M_\Earth$, $17M_\Earth=1M_{\rm N}$, and $54M_\Earth$, respectively, assuming Neptune's composition and accounting for the gravitational compression of the core. For each planet size, we display atmosphere masses ranging roughly from 1\% to 50\% of the planet's mass.\label{fig:teff}}
\end{figure}

\begin{figure}[tbhp]
	\includegraphics[width=\columnwidth]{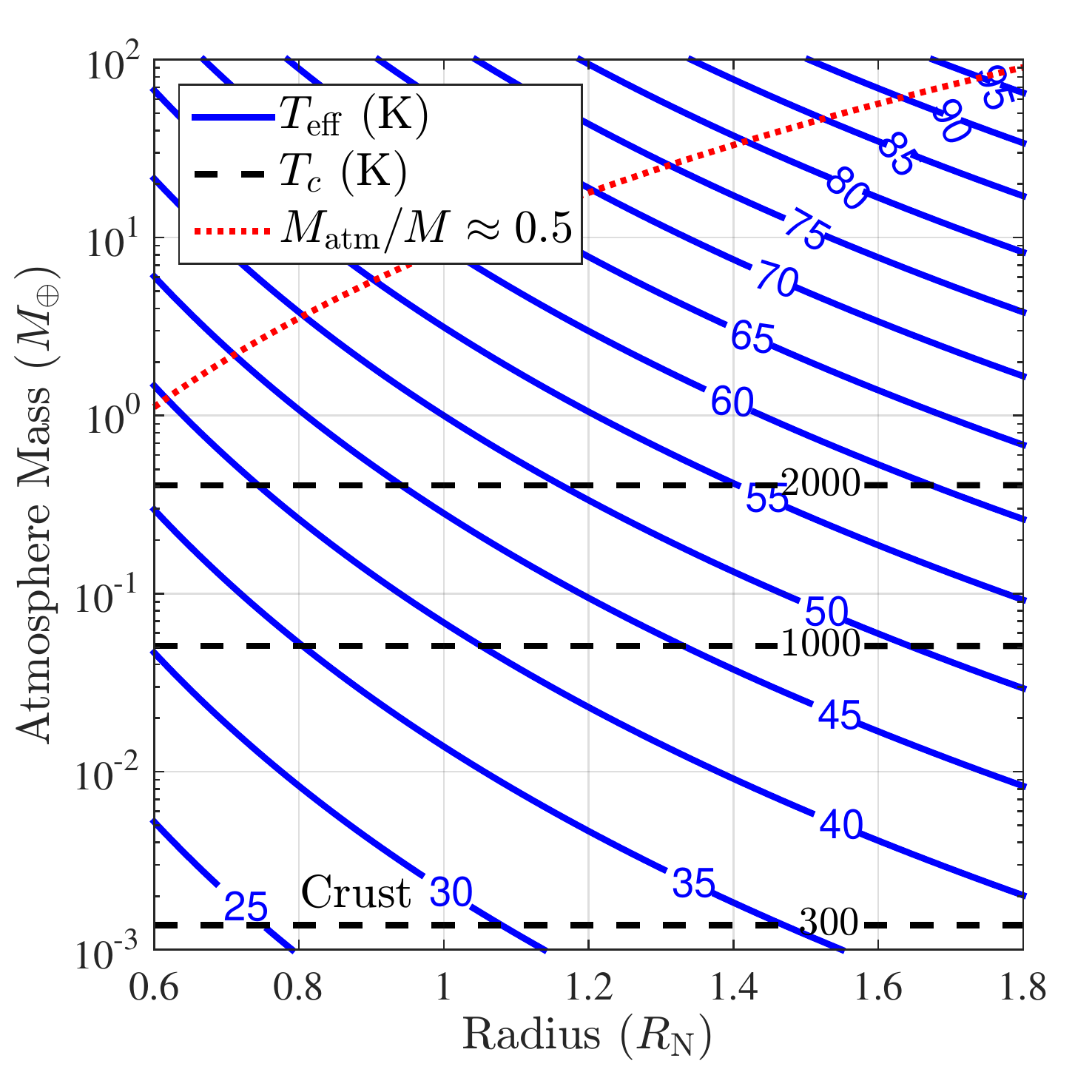}
	\caption{Contours of the effective temperature (solid blue lines) from Equation \eqref{eq:tmp_rad} and of the temperature at the atmosphere-core boundary (dashed black lines) according to Equation \eqref{eq:tc_rad} at an age of 4.5 Gyr, as a function of the planet's radius and atmosphere mass. A solid crust forms below the bottom dashed black line, and the cooling is no longer dominated by the atmosphere. An upper mass fraction limit of $f\equiv M_{\rm atm}/M\approx 0.5$ (dotted red line) is also provided, assuming Neptune's composition and accounting for the gravitational compression of the core. \label{fig:teff_cont}}
\end{figure}

In Figures \ref{fig:teff}  and \ref{fig:teff_cont} we demonstrate the relation between the planet's effective temperature, its size and its atmosphere's mass. Due to the weak dependence implicit in Figure \ref{fig:teff} and Equation \eqref{eq:tmp_rad}, an accurate measurement of the temperature is required for a tight constraint on the atmosphere mass. Such accuracy is possible at the Wien tail of the spectrum, where the dependence of the Planck function on the temperature is exponential.  

Similarly, using Equations \eqref{eq:tmp_ratio} and \eqref{eq:tmp_rad} we estimate the temperature at the atmosphere-core boundary at the age of the solar system
\begin{equation}\label{eq:tc_rad}
T_c\approx 2700\textrm{ K}\left(\frac{M_{\rm atm}}{M_\Earth}\right)^{1/3}.
\end{equation}
If this temperature is below the melting point of ice, then an insulating solid crust forms and the cooling is no longer mediated only by the atmosphere. The phase diagram of ice is given by \citet{Lobban1998} and \citet{Redmer2011}, with the pressure at the boundary determined by
\begin{equation}\label{eq:pressure}
P=\frac{GMM_{\rm atm}}{4\pi R^4}\approx 0.1\textrm{ Mbar}\left(\frac{M_{\rm atm}}{M_\Earth}\right).
\end{equation}
By comparing the contours of $T_c$ with Equation \eqref{eq:pressure} and the phase diagrams of \citet{Lobban1998} and \citet{Redmer2011}, we conclude that a solid crust forms only when the melting curve of ice flattens, close to 300 K, reached by atmospheres lighter than $\approx 10^{-3}M_\Earth$, as seen in Equation \eqref{eq:tc_rad} and Figure \ref{fig:teff_cont}.

\begin{figure}[tbhp]
	\includegraphics[width=\columnwidth]{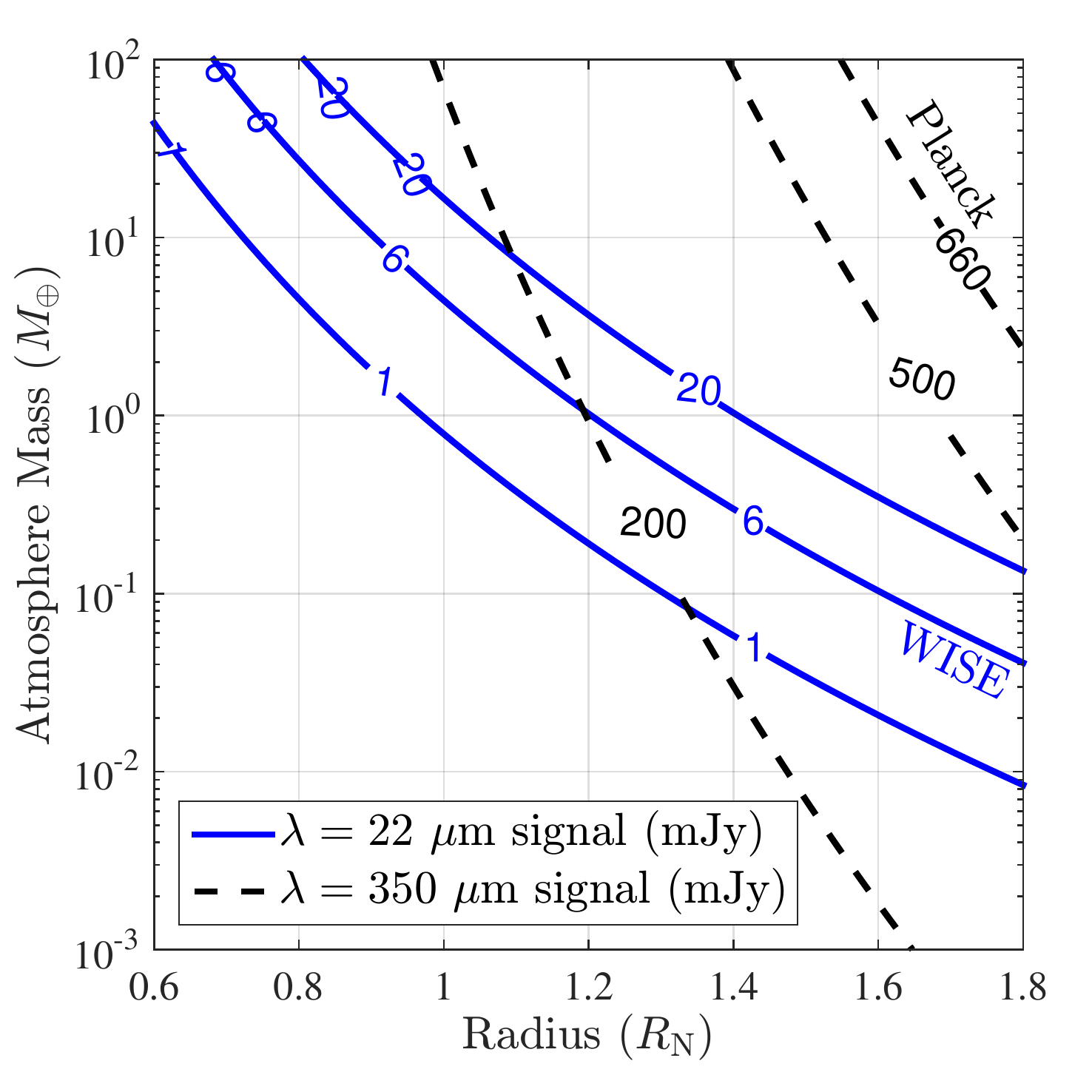}
	\caption{Contours of the observable flux density at wavelengths of $22\textrm{ }\mu\textrm{m}$ (solid blue lines) and $350\textrm{ }\mu\textrm{m}$ (dashed black lines) of 4.5 Gyr old planets at a distance of 700 AU, as a function of their radius and atmosphere mass. The $22\textrm{ }\mu\textrm{m}$ sensitivity of WISE is 6 mJy \citep{Wright2010}, corresponding to the middle solid blue line, while the $350\textrm{ }\mu\textrm{m}$ sensitivity of Planck is 660 mJy \citep{Planck2013,Planck2015}, corresponding to the top dashed black line. \label{fig:detection}}
\end{figure}

The peak emission of a blackbody at $T_{\rm eff}\approx50\textrm{ K}$ is at a wavelength of $\lambda\approx60\textrm{ }\mu\textrm{m}$. In Figure \ref{fig:detection} we present the observable flux density at similar wavelengths using the temperatures calculated by Equation \eqref{eq:tmp_rad} and Figure \ref{fig:teff_cont}, assuming a distance of 700 AU. We also compare these flux densities to the sensitivities of WISE \citep{Wright2010} and Planck \citep{Planck2013,Planck2015}. As seen in Figure \ref{fig:detection}, a null detection by WISE can (marginally) constrain the size and atmosphere mass of Planet Nine (below the middle solid blue line), while a null detection by Planck is less informative.

\section{Conclusions}\label{sec:conclusions}

Recent studies suggest the existence of an undiscovered Neptune-size planet in the outer solar system. The formation of one or several such planets is not unnatural within the context of standard formation theories of the solar system \citep{BromleyKenyon2014,KenyonBromley2015}. The cooling radiation of Neptunes on such distant orbits surpasses the flux they receive (and re-emit) from the Sun.

In this work we modeled these planets using a simple analytical two-layer model that consists of a rocky/icy core and a hydrogen/helium envelope. We examined the evolution of the planets over time and derived scaling laws relating their effective temperature to their size and composition, enabling an interpretation of a future detection (or null detection).

Specifically, we demonstrated that the effective temperature at a given age, $T_{\rm eff}$, depends on the mass of the atmosphere, $M_{\rm atm}$, since it mediates the planet's cooling. Explicitly, we found the relation $T_{\rm eff}\propto M_{\rm atm}^{1/12}$.
Despite this weak relation, an accurate measurement of the effective temperature is possible in the Wien tail of the spectrum, and it can constrain the mass of the atmosphere at least to within a factor of a few.  Even such a rough estimate can be useful in distinguishing between gas rich and mainly rocky/icy planets, providing a clue to their formation scenario. From Figure \ref{fig:teff_cont} we find that planets with atmospheres lighter than $\sim 10^{-3}M_\Earth$ develop a solid insulating crust with a different simple two-layer model, of a convective core and a conducting crust, being adequate in this case \citep[see, e.g.,][]{Stevenson1983}. 

The approximate analytical scaling relations we present here should be supplemented by more elaborate numerical models to constrain the mass and composition of such planets, if discovered. Nonetheless, this study provides an intuitive demonstration of what we are able to learn from future observations. In addition, we demonstrate that a null detection by recent all-sky surveys can constrain the size and composition of Planet Nine.

\acknowledgements
This research was supported in part by ISF, ISA and iCore grants, and by NSF grant AST-1312034. 

\appendix

\section{Convective Atmosphere}\label{sec:convective}

In Section \ref{sec:cooling} we assumed that the atmosphere is radiative and has a constant opacity. Here we treat the atmosphere more generally, allowing for a variable opacity and convection.

Using Equation \eqref{eq:tmp_prof} we find that an adiabatic temperature profile scales with optical depth $\tau=\int{\kappa\rho\rm{d}x}$ as $T\propto\tau^{(\gamma-1)/\gamma}$ if the opacity is constant. Convective instability sets in if the radiative profile $T\propto\tau^{1/4}$ is steeper than the adiabatic one, i.e. $\gamma<4/3$. Therefore, molecular hydrogen ($\gamma=7/5$) is indeed stable against convection and if the opacity is constant, the atmosphere is radiative.

In the case of a variable opacity, which we model with a power law $\kappa\propto\rho^a T^b$, an adiabatic profile scales as
\begin{equation}\label{eq:adiabatic_kappa}
T\propto\tau^{(\gamma-1)/(\gamma+a+b\gamma-b)}.
\end{equation}
For the diatomic $\gamma=7/5$ Equation \eqref{eq:adiabatic_kappa} shows that even a relatively modest increase of the opacity with depth (temperature or density) leads to convection. For example, if $b=0$ then $a>0.2$ is convectively unstable, while if $a=0$ then $b>0.5$ suffices. In this case, the atmosphere's profile will be adiabatic, given by Equation \eqref{eq:adiabatic_kappa}, and flatter in comparison with a radiative one.

We conclude that the temperature ratio given by Equation \eqref{eq:tmp_ratio}, $T_c/T_{\rm eff}\propto\tau_{\rm atm}^{1/4}$, provides an upper boundary to the steepness of the temperature profile and consequently our scaling of $T_{\rm eff}\propto M_{\rm atm}^{1/12}$ is also an upper bound to the dependence on $M_{\rm atm}$. In the relevant regime, the opacity is almost constant \citep{Alexander1989,BellLin1994} so even if the profile is convective, it is not very different from the radiative, $T\propto\tau^{1/4}$, chosen for simplicity above.

\bibliography{bib}{}
\bibliographystyle{apj}
  
\end{document}